\documentclass[prl,showpacs,twocolumn,superscriptaddress,floatfix]{revtex4}
\usepackage{graphicx}
\usepackage{dcolumn,float}
\usepackage{amsmath,amssymb}
\usepackage{subfigure}

\newcommand{\be}{\begin{eqnarray}}

\newcommand{\ee}{\end{eqnarray}}

\newcommand{\qq}{\begin{eqnarray}}
\newcommand{\qqq}{\end{eqnarray}}

\newcommand{\re}[1]{(\ref{#1})}

\newcommand {\dis}{\displaystyle}

\newcommand{\beg}{\begin{equation}}
\newcommand{\en}{\end{equation}}

\newcommand{\eps}{\epsilon}
\newcommand{\lam}{\lambda}

\begin{document}

\title{BCS theory for finite size superconductors}

\author{Antonio M. Garc\'{\i}a-Garc\'{\i}a}
\affiliation{Physics Department, Princeton University, Princeton, New Jersey 08544, USA}
\author{Juan Diego Urbina}
\affiliation{Institut F\"ur Theoretische Physik, Universit\"at Regensburg, 93040 Regensburg, Germany}
\affiliation{Department of Physics, Universidad Nacional de Colombia, Cll45 Cra 30, Bogota, Colombia}
\author{Emil A. Yuzbashyan}
\affiliation{Center for Materials Theory, Rutgers University, Piscataway, New Jersey 08854, USA}
\author{Klaus Richter}
\affiliation{Institut F\"ur Theoretische Physik, Universit\"at Regensburg, 93047 Regensburg, Germany}
\author{Boris L. Altshuler}
\affiliation{Physics Department, Columbia University, 538 West 120th Street, New York, NY 10027, USA}
\affiliation{NEC-Laboratories America, Inc., 4 Independence Way, Princeton, NJ 085540, USA}

\begin{abstract}
We study finite size effects in
superconducting metallic grains and determine the BCS order parameter and the low energy
excitation spectrum in terms of size, and shape of the grain.
Our approach combines the BCS self-consistency condition,   a semiclassical expansion for the spectral density and  interaction matrix elements, and corrections to the BCS mean-field. In chaotic grains mesoscopic fluctuations of the matrix elements lead to a smooth dependence of the order parameter on   the excitation energy.  In the integrable case we observe shell effects when e.g. a small change in the electron number leads to large changes
in the energy gap.
\end{abstract}


\maketitle

\newcommand{\bb}{\boldsymbol{\beta}}
\newcommand{\ba}{\boldsymbol{\alpha}}

Since experiments by Ralph, Black, and Tinkham \cite{tinkham} on Al nanograins
  in mid nineties, there has been considerable interest
in the theory of ultrasmall superconductors (see \cite{parmenter,muhl} for earlier studies).  In particular, finite-size corrections to the predictions of the Bardeen, Cooper, and Schriffer (BCS) theory for bulk superconductors \cite{BCS} have been studied [\citealp{ML}--\citealp{yuzbashyan}] within the exactly solvable Richardson model \cite{richardson}. Pairing in specific potentials, such as a harmonic oscillator potential \cite{heiselberg} and a rectangular box, \cite{peeters,fomin} and mesoscopic fluctuations of the energy gap \cite{shuck,leboeuf} have been explored as well. Nevertheless, a comprehensive theoretical description of the combined effect of discrete energy spectrum and fluctuating interaction matrix elements has not yet emerged.
We note that the Richardson model
alone cannot  provide such a description as  it does not allow for mesoscopic fluctuations of the matrix elements.

In the present paper we develop a framework based on the BCS theory and semiclassical
techniques that permits a systematic analytical evaluation of the low energy spectral properties of superconducting nanograins in terms of their size and shape. Leading finite size corrections to the BCS mean-field can also be taken into account in our approach. Our main results are as follows. For chaotic grains, we show that the order
parameter is energy dependent. The energy dependence is universal, i.e. its functional form is the same for
all chaotic grains. The matrix elements are responsible for most of the deviation from  the bulk limit. 
In integrable grains,  we find that the superconducting gap is strongly sensitive to shell effects, namely, a
small modification of the grain size or number of electrons can substantially affect its value. 

We start with the BCS Hamiltonian, $H=\sum_{n\sigma}\eps_n c_{n\sigma}^\dagger c_{n\sigma}-\sum_{n,n'}I_{n,n'} c_{n\uparrow}^\dagger c_{n\downarrow}^\dagger c_{n'\downarrow}c_{n'\uparrow},$
where $c_{n\sigma}$ annihilates an electron of spin $\sigma$  in state $n$,
\be
I_{n,n'} = I(\eps_{n},\eps_{n'})=\lam V \delta \int \psi^2_n(r)\psi_{n'}^{2}(r)dV
\ee
are matrix elements of a short-range electron-electron interaction, $\lam$ is the BCS coupling constant, and
$\psi_{n}$ and $\eps_n$ are  eigenstates and eigenvalues  of the one-body mean-field Hamiltonian of a free particle of mass $m$ in a clean grain of volume $V$. Eigenvalues $\eps_n$ are measured from the Fermi level $\eps_F$ and the
 mean level spacing $\delta = 1/\nu_{\rm\small TF}(0)$, where $\nu_{{\small \rm TF}}(0)=2
\frac{V}{4 \pi^{2}}\left(\frac{2 m}{\hbar^{2}}\right)^{3/2}\sqrt{\eps_{F}}$ is the spectral density at the Fermi level
 in the Thomas-Fermi approximation.

Our general strategy can be summarized as follows:
a) use semiclassical techniques to compute the spectral density  $\nu(\eps)=\sum_{n} \delta(\eps-\eps_n)$ and
$I(\eps,\eps')$  as series in a small parameter $1/ k_F L$, where $k_{F}$ is the Fermi wavevector and $L \simeq V^{1/3}$ is the size of the grain
b) solve the BCS gap equation in orders in $1/k_FL$ c) evaluate the low energy spectral properties of the grain
such as the
energy gap, excitation energies, and Matveev-Larkin parameter \cite{ML} including finite size corrections to the
BCS mean-field.  
The results thus obtained are strictly valid in the region, $k_F L \gg 1$ (limit of validity of the semiclassical approximation), $\delta/\Delta_0 < 1$ (limit of validity of the BCS theory), and  $l\gg\xi \gg L$ (condition of quantum coherence) where $\xi=\hbar v_F/\Delta_0$ is the superconducting coherence length, $v_F$ is
 the Fermi velocity, $l$ is the coherence length of the single particle problem and $\Delta_0$ is the bulk gap.
We note that in Al grains \cite{albreak} $\xi \approx 1600 nm$ and $l > 10000nm$ for temperatures $T \leq 4K$. Therefore the region $l\gg\xi \gg L$ is accessible to experiments. 

Since the matrix elements $I(\eps,\eps')$ are energy dependent the BCS order parameter $\Delta(\eps)$ also depends
on energy. The self-consistency equation for $\Delta(\eps)$ reads
\begin{equation}
\Delta(\eps)=\int_{-E_D}^{E_D}\frac{ \Delta(\eps')I(\eps,\eps') }{ 2\sqrt{ {\eps'}^2+\Delta(\eps')^2} }\nu(\eps')d\eps',
\label{gap}
\end{equation}
 where $E_{D}$ is the Debye energy.
In the limit $V \to \infty$, the spectral density in the $2E_D$ energy window near the Fermi level can be taken to
be energy independent and given by the Thomas-Fermi approximation,
$\nu(\eps)=\nu_{{\small \rm TF}}(0)$,  matrix elements
are also energy independent, $I(\eps,\eps')= \lam \delta$, and the gap is equal to its bulk value,
$\Delta_{0}= 2E_{D}{\rm e}^{-\frac{1}{\lambda}}$.
As the volume of the grain decreases the mean level spacing  increases and eventually both
 $\nu(\eps)$ and $I(\eps,\eps')$ deviate from the bulk limit.

{\it Semiclassical evaluation of $\nu(\eps)$.} The spectral density in a 3$d$ grain,
\begin{equation}
\label{ge}
\nu(\eps')\simeq \nu_{{\small\rm TF}}(0)\left[1+\bar{g}(0)+\tilde{g}_{l}(\eps')\right]
\end{equation}
consists of a monotonous part,
$\bar{g}(0)=
\pm \frac{{S}\pi}{4k_{F}V}+\frac{2{ C}}{k_{F}^{2}V}$ and an oscillatory contribution $\tilde{g}_{l}(\eps')$. Here
$S$ and $C$ denote the surface area and mean curvature of the grain, respectively, and
upper/lower signs stand for Neumann/Dirichlet boundary conditions. The oscillatory contribution, to leading order, is given by the Gutzwiller trace formula \cite{gut,baduri},
\beg
\begin{array}{l}
\dis \tilde{g}_{l}(\eps')= \Re
\frac{2\pi}{k_{F}^{2}V}\sum_{p}^{l}A_{p}e^{\dis i \left(k_F L_p+\beta_p+\frac{\eps' k_F L_p}{ 2\eps_F}\right)}.
\end{array}
\label{dos}
\en
where both the amplitude $A_p$ and the topological index $\beta_p$ depend on classical quantities only \cite{baduri}.
The summation is over a set of classical periodic orbits $p$ of length $L_p$.  For isolated grains Dirichlet is the most natural choice, but we also include Neumann to illustrate the dependence of our results on boundary conditions.
Only orbits shorter than the quantum coherence length $l$ of the single-particle problem are included.  This effectively accounts for inelastic scattering and other factors  that destroy quantum coherence. 
Here we focus on the limit $l\gg\xi$,  the case $l \sim \xi$ will be discussed  elsewhere \cite{uslong}. 
In Eq.~\re{dos}  classical actions $\hbar k(\eps') L_{p}$ are expanded as $k(\eps') \approx k_{F}+\eps'k_{F}/2\eps_{F}$.  The amplitude $A_{p}$ 
increases by a factor ${(k_FL)}^{1/2} \gg 1$ for each of the symmetry axes of the grain. 

{\it Semiclassical evaluation of $I(\eps,\eps')$.}
For integrable systems $I(\eps,\eps')$ depends on  details of the system. In a rectangular box it is
simply $I(\eps,\eps')= \lam\delta$ but in most other geometries an explicit expression in terms of classical
quantities is not available.
In the chaotic case the situation is different. As a result of the quantum ergodicity theorem \cite{erg} it is well justified to assume that for systems 
with time reversal symmetry (the only ones addressed in this paper),
$\psi^{2}_{n}(\vec{r})=\frac{1}{V}(1+ O(1/k_FL))$.   
  In order to explicitly determine deviations from the bulk limit we replace  $\psi^{2}_{n}(\vec{r})$ in $I_{n,n'}$ with $\langle \psi^{2}(\vec{r})\rangle_{\eps_{n}}$,  where $\langle \ldots \rangle_{\eps}$ stands for an energy average around $\eps$.  The single-particle probability density is thus
effectively averaged over a small energy window resembling the effect of a finite coherence length. 

  Substituting $\langle \psi^{2}(\vec{r}) \rangle _{\eps}$ into $I_{n,n'}$, we obtain
\begin{equation}
\label{Is}
I(\eps,\eps')=\frac{\lambda}{V}\left[1-\left(\frac{ S\pi}{4k_{F}V}\right)^{2}+\bar{I}(\eps_{F},\eps,\eps')\right],
\end{equation}
where
\begin{equation}
\bar{I}(\eps_{F},\eps,\eps')=\bar{I}^{{\rm short}}(\eps_F)+\bar{I}^{{\rm long}}(\eps_F, \eps-\eps')
\end{equation}
can be split into two parts coming from short and long orbits.
Short orbits  involve a single reflection at the grain boundary and result in a monotonous contribution
\begin{equation}
\label{Ishort}
\bar{I}^{{\rm short}}(\eps_{F})=\frac{\pi{\cal S}}{4k_{F}V},
\end{equation}
while the contribution of long orbits depends on the energy difference $\eps-\eps'$
\begin{equation}
\label{Ilong}
\bar{I}^{{\rm long}}(\eps_{F},\eps-\eps')=
\frac{1}{V}\Pi_{l}\left(\frac{\eps-\eps'}{\eps_{F}}\right),
\end{equation}
with $\Pi_{l}(w)=\int\sum_{\gamma(r)}^{l}D_{\gamma}^{2}\cos{\left[wk_{F}L_{\gamma}\right]}dr$, where the sum is over all non-zero classical paths (not periodic orbits) $\gamma(r)$ starting and ending at a given point $r$ inside the grain \cite{KJDU1} and the 
 amplitude $D_{\gamma}$ is defined in Refs.~\cite{gut,uslong,KJDU1}. The integral stands for an average over all points $r$ inside the grain.
The explicit evaluation of $\Pi_{l}(w)$ for a given geometry requires in principle the knowledge of all classical paths $L_\gamma$ up to length $l$. 
However, for $l \gg L$,
one can use a {\it sum rule} for classical closed orbits \cite{martinp} to obtain
\begin{equation}
\label{pi}
\Pi_{l}(w)=\left(\frac{2 \pi}{k_{F}}\right)^{2}\frac{\sin(w k_F l)}{w k_{F}}.
\end{equation}

\begin{figure}[ht]
\includegraphics[width=0.9\columnwidth,height=0.6\columnwidth,clip,angle=0]{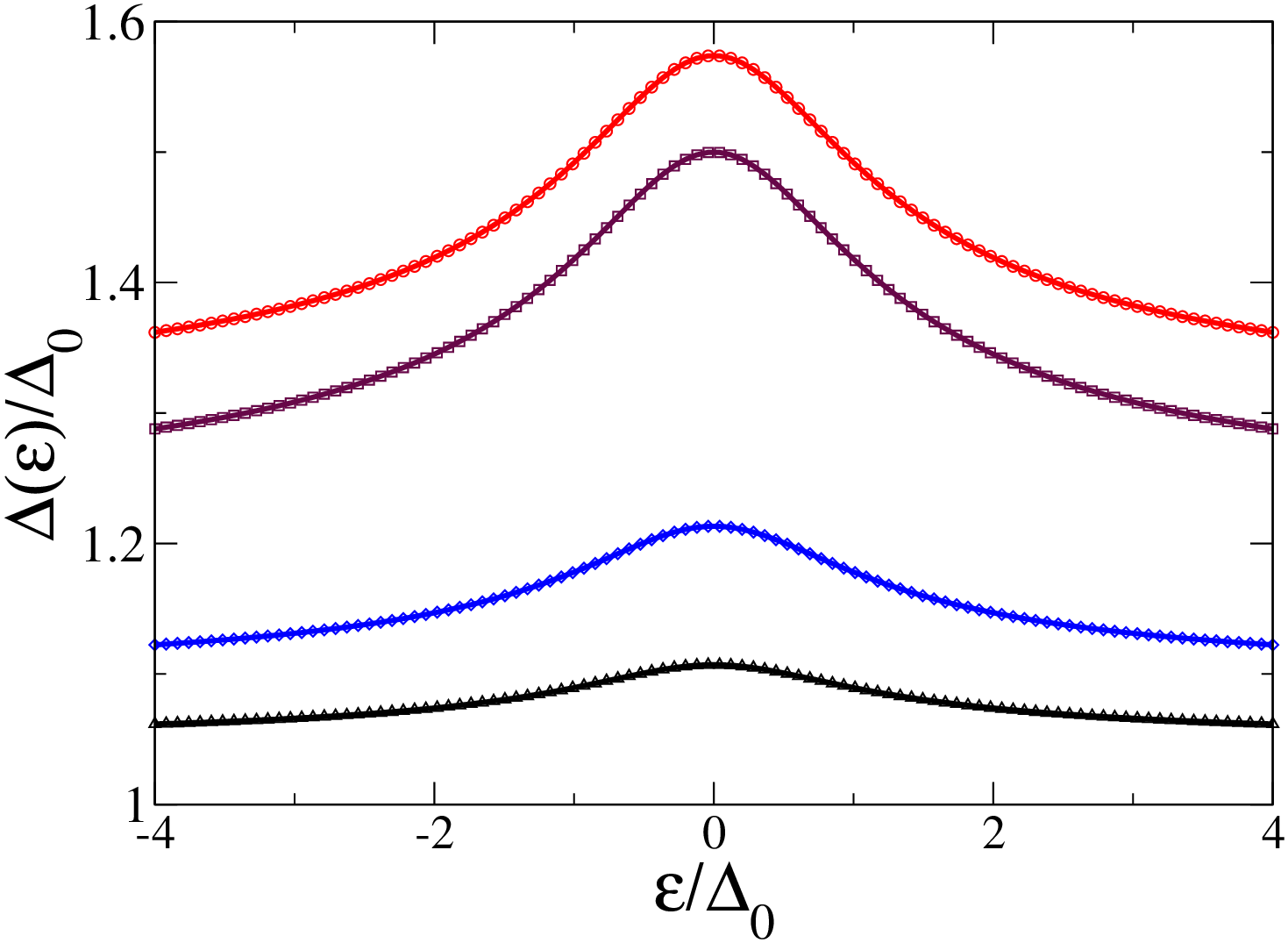}
\vspace{3mm}
\caption{Superconducting order parameter $\Delta(\eps)$ in units of the bulk gap $\Delta_0$ for chaotic Al grains 
($k_F = 17.5 nm^{-1}, \delta = 7279/N, \Delta_0 \approx 0.24 meV$) as a function of energy $\eps$ counted from the Fermi level. Different curves correspond to grain sizes (top to bottom)  
$L=6nm, k_F L = 105, \delta/\Delta_0 = 0.77)$ (Dirichlet and Neumann boundary conditions), $L= 8nm, k_F L = 140, \delta/\Delta_0 = 0.32$ (Dirichlet), and  $L= 10nm, k_F L = 175, \delta/\Delta_0 = 0.08$  (Dirichlet).  The leading contribution comes from the energy dependent matrix elements $I(\eps,\eps')$ given by Eq. (\ref{Is}).} 
\label{gap3D}
\end{figure}

{\it Solution of the gap equation.}
First, let us consider chaotic grains. Here we present only the final answer for the 3$d$ case
deferring a more detailed account, including the 2$d$ case, to Ref.~\cite{uslong}.
Writing the gap function $\Delta(\eps)$ formally as a series in $1/k_{F}L$,
\begin{eqnarray}
\label{Delta3D}
\Delta(\eps)
 &=&\Delta_0\left[1+f^{(1)}+ f^{(2)}+ f^{(3)}(\eps)\right],
\end{eqnarray}
substituting it into Eq.~\re{gap}, and using   the
above expressions for the density of states and interaction matrix elements, we derive
\be
\label{f1}
f^{(1)}=\frac{1\pm 1}{\lambda}\frac{\pi {\cal S}}{4k_{F}V},
\ee
where $\pm$ stands for Neumann ($+$) and Dirichlet ($-$) boundary conditions.
Note that   to leading order the combined effect of the interaction matrix elements and the density of states have very different consequences on the gap, depending on the kind of boundary conditions.  For Dirichlet  the leading  finite size corrections to the gap vanishes.

The second order ($1/(k_FL)^2$) correction reads
\begin{equation}
\label{f2}
\lam f^{(2)}=  \frac{2{\cal C}}{k_{F}^{2}{V}}+
2\left(\mp 1+ \frac{1\pm 1}{\lambda}\right)\left(\frac{\pi {\cal S}}{4 k_{F}V}\right)^{2}+\tilde{g}(0),  \nonumber
\end{equation}
where,
\beg
\tilde{g}(0) =
\frac{2\pi}{k_{F}^{2}V}\sum_{p}^{l}A_{p}W(L_p/\xi)\cos(k_{F}L_{p}+\beta_p)
\label{g0}
\en
and,
\beg
W(L_p/\xi) = \frac{\lambda}{2} \int_{- \infty}^{\infty}dt \frac{\cos(L_p t/\xi)}{\sqrt{1+t^2}}
\label{W}
\en
exponentially suppresses periodic orbits longer than $\xi$.

The third order correction (included in the definition of $\delta$) is energy dependent,
\begin{eqnarray}
\label{fx}
f^{(3)}(\eps)&=&\frac{\pi \lambda
\delta}{\Delta_0}\left[\frac{\Delta_0}{\sqrt{\eps^{2}+\Delta_0^2}}+\frac{\pi}{4}\right].
\end{eqnarray}
Note that a)  $\delta/\Delta_0 \ll 1$ is an additional expansion parameter, therefore the contribution \re{fx} can be comparable to
lower orders in the expansion in $1/k_FL$ and
 b) the order parameter $\Delta(\eps)$ has a maximum at the Fermi energy ($\eps=0$) and slowly
decreases on an energy scale $\eps\sim\Delta_0$ as one moves away from the Fermi level.  One can also show that mesoscopic corrections given by Eqs.~(\ref{f1},\ref{f2}) and \re{fx} always enhance $\Delta(0)$ as compared to the bulk value $\Delta_0$. Fig.~\ref{gap3D} shows the gap function $\Delta(\eps)$  for Al grains of different sizes
$L$, where we used (see \cite{tinkham})  $k_F \approx 17.5 \mbox{nm}^{-1}$, $\lambda \approx 0.18$, and $\delta \approx 7279/N \mbox{ meV}$,  where $N$ the number of particles.

Several remarks are in order: a) the smoothing of the spectral density energy dependence in Eq.~\re{g0} caused by a cutoff function $W$ is a superconductivity effect not related to the destruction of quantum coherence, b) the energy dependence of the gap is universal in the sense that it does not depend on
specific grain details, c) the matrix elements $I(\eps,\eps')$ play a crucial role, e.g.
  they are responsible for most of the deviation from  the bulk limit in Fig.~\ref{gap3D},
d) the requirement $\xi \gg L$ used to derive Eq.~\re{pi} is well justified for nanograins since $L \sim 10 \mbox{nm}$, while   $\xi \sim  10^4 \mbox{nm}$.

We now turn to the integrable case.
Probably the simplest   example is that of a rectangular
box, since in this case the interaction matrix elements are simply $I(\eps,\eps') = \lambda\delta$.
The calculation is simplified as now the order parameter is energy independent. We have
\be
\label{int1}
\Delta=\Delta_{0}\left[1+f^{(1)}+f^{(3/2)}+f^{(2)}\right],
\ee
where $f^{(n)}\propto {\left(k_F L \right)}^{-n}\lambda^{-1}$.
We obtain
\beg
\begin{array}{l}
\dis
\lam f^{(1)}= \bar{g}(0)+\tilde{g}^{(1)}(0),\\
\\
\dis \lam f^{(3/2)}= \sum_{i,j\neq i}\tilde{g}^{(3/2)}_{i,j }(0),\\
\\
\dis \lam f^{(2)}= \sum_{i}\tilde{g}^{(2)}_{i}(0)+f^{(1)}[f^{(1)} -\bar{g}(0)],\\
\end{array}
\en
where ${\tilde g^{(k)}} \propto \left(k_F L \right)^{-k}$ denotes the oscillating part of the spectral density and
indexes $i$ and $j$ take values 1, 2, and 3 in three dimensions.   Explicit expressions for $\tilde{g}^{(k)}$,  $\tilde{g}^{(k)}_i$, and $\tilde{g}^{(k)}_{i,j}$ in terms of periodic orbits for a rectangular box can be found in
Ref.~\cite{baduri} (the cutoff function in   our case is given by Eq.~\re{W}).
We note  that: a) Eq.(\ref{int1}) is also obtained by expanding the standard expression of the bulk gap
$\Delta=2E_D\exp(-\nu_{\small\rm TF}(0)/\nu(0)\lam)$
in powers of  $(k_F L \lambda)^{-1}$ with $\nu(0)$ given by Eq.(\ref{ge}). 
b) unlike the chaotic case,  the leading smooth correction to the bulk limit does not vanish for any boundary condition, c) smooth and
oscillating corrections are of comparable magnitudes.
\begin{figure}
\includegraphics[width=4.2cm,height=0.55\columnwidth,clip,angle=0]{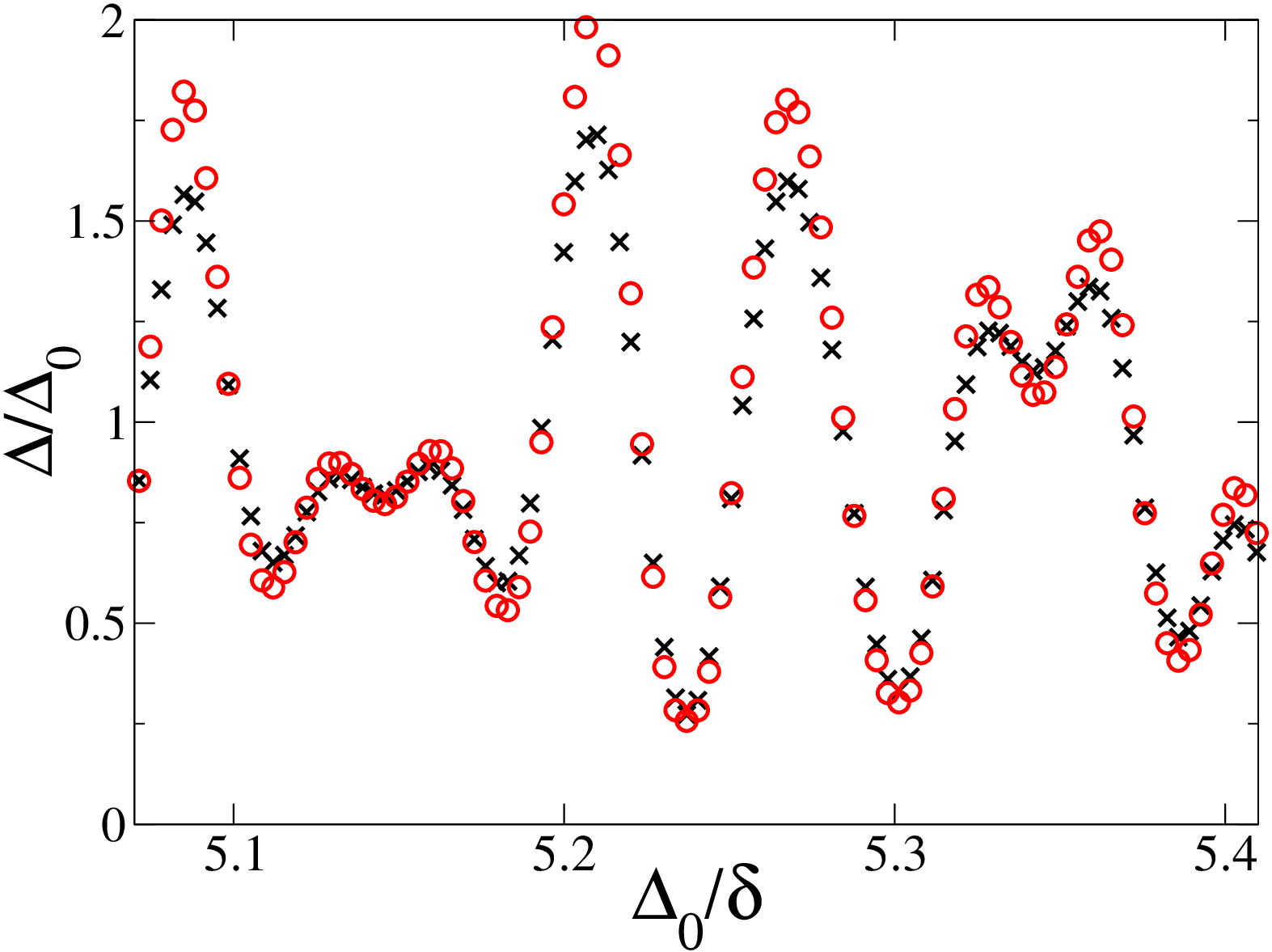}
\includegraphics[width=4.2cm,height=0.55\columnwidth,clip,angle=0]{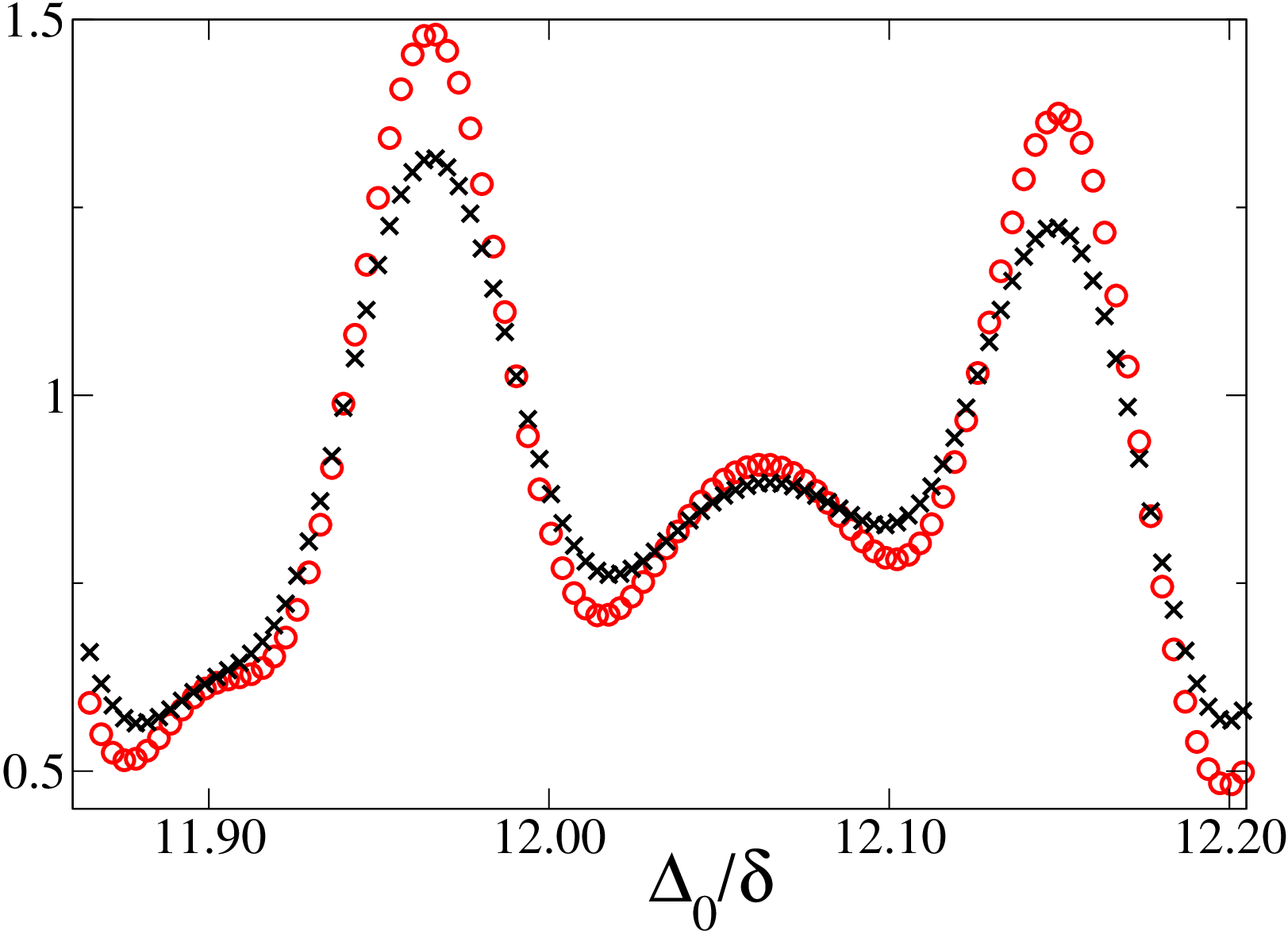}
\vspace{3mm}
\caption{
Superconducting order parameter $\Delta$ in units of the bulk gap $\Delta_{0}$ for a cubic Al grain
as a function of the ratio $\Delta_0/\delta$, where $\delta$ is the mean level spacing.
Black crosses correspond to the exact numerical solution of the gap equation (\ref{gap}), while the red circles represent the semiclassical analytical expression (\ref{int1}).}
\label{gapN}
\end{figure}

{\it Shell effects and fluctuations.}
Motivated by previous studies for other fermionic systems  such as nuclei and atomic clusters (see e.g. Ref.~\cite{metal}), we investigate shell effects in metallic nanograins. In particular,  we are interested in the fluctuations of the BCS gap with the number of electrons on the grain.
As an illustration let us consider a cubic geometry.
To determine the gap,  we solve the gap equation (\ref{gap}) numerically and  determine the Fermi energy for a given
 number of electrons $N$ by inverting
 the relation $2\int^{\eps_F}\nu(\eps)d\eps = N$.  We find a good agreement between numerical results and
the semiclassical expansion (\ref{int1}), see Fig.~\ref{gapN}.
We also observe that a slight modification of the grain size (or equivalently the number of electrons
$N$ or the mean level spacing $\delta$) can result in substantial changes in the value of the gap, see Fig.~\ref{gapN}.
The typical magnitude of
fluctuations of the gap, $\frac{\tilde \Delta }{\Delta_0} \approx \sqrt{\frac{\pi\delta}{4\Delta_0}}$ \cite{leboeuf}
is consistent with our results (see Fig. 2). 

{\it Low energy excitations.} Having solved the gap equation \re{gap}, one can evaluate  low energy properties of the grain taking into account finite size corrections to the BCS mean-field approximation. For example,
the energy cost for breaking a Cooper pair in an isolated grain is \cite{ambe},
\beg
\Delta E=2\Delta(0)-\delta,
\label{engap}
\en
where $\Delta(0)$ is the solution of equation \re{gap} taken at the Fermi energy and is given by Eqs.~\re{Delta3D} and \re{int1} for chaotic and rectangular shapes, respectively. We note that the correction
to the mean-field ($-\delta$) has been evaluated \cite{yuzbashyan} for constant interaction matrix elements. Nevertheless, since
the deviation of matrix elements from a constant energy independent value is itself of order $(k_FL)^{-1}$,
Eq.~\re{engap} is accurate up to terms of order $(\delta/\Delta_0)(k_FL)^{-1}$,   which are negligible as compared
to the ones we kept in Eqs.~(\ref{engap}), (\ref{Delta3D}), and \re{int1}.

Similarly,  the Matveev-Larkin parity parameter \cite{ML} reads
$
\Delta_p \equiv E_{2N+1} - \frac12\big(E_{2N}+E_{2N+2}\big)=\Delta(0) - \frac{\delta}{2},
$
where  $E_N$ is the ground state energy for a superconducting grain with $N$ electrons. Quasiparticle energies
are $\sqrt{\eps^2+\Delta(\eps)^2}$ plus  corrections to mean-field, which can be determined
using the approach of Ref.~\cite{yuzbashyan}.

We see that finite size corrections to the BCS mean-field approximation are comparable to the energy dependent  correction (\ref{Delta3D}) obtained within mean-field, but have an opposite sign. We also note that our approach
of expanding around the bulk BCS ground state is applicable only when $\delta\ll\Delta_0$, i.e. when corrections
to the BCS mean-field approximation are small \cite{ander2}.

To conclude,  we have determined the low energy excitation spectrum for small superconducting grains
 as a function of their size and shape  by combining the BCS mean-field, semiclassical techniques and
 leading corrections to the mean-field. For chaotic grains the non-trivial energy dependence of the interaction matrix elements
 leads to a universal smooth dependence \re{fx} of the  gap function on excitation energy.
 In the integrable case we found that  small changes in the number of electrons   can substantially modify
 the superconducting gap.

A.M.G. thanks Jorge Dukelsky for fruitful conversations. K.R. and J.D.U. acknowledge conversations with Jens Siewert and financial support from the Deutsche Forschungsgemeinschaft (GRK 638). E.A.Y. was  supported by Alfred P. Sloan Research Fellowship and NSF award NSF-DMR-0547769.


\end{document}